\documentclass[conference,letterpaper]{IEEEtran}
\usepackage{graphicx}
\usepackage{wrapfig}
\usepackage{xcolor}
\usepackage[utf8]{inputenc} 
\usepackage[T1]{fontenc}
\usepackage{url}
\usepackage{ifthen}
\usepackage{cite}
\usepackage[cmex10]{amsmath} 
\usepackage[normalem]{ulem}

\begin{document}
\title{Memory Access Vectors: Improving Sampling Fidelity for CPU Performance Simulations}

\author{%
  \IEEEauthorblockN{Sriyash Caculo, Mahesh Madhav, Jeff Baxter}
  \IEEEauthorblockA{Ampere Computing, Portland, OR \\
                    Email: \{ycaculo, mahesh, jbaxter\}@amperecomputing.com}

}

\maketitle


\begin{abstract}
Accurate performance projection of large-scale benchmarks is essential for CPU architects to evaluate and optimize future processor designs. SimPoint sampling, which uses Basic Block Vectors (BBVs), is a widely adopted technique to reduce simulation time by selecting representative program phases. However, BBVs often fail to capture the behavior of applications with extensive array-indirect memory accesses, leading to inaccurate projections. In particular, the 523.xalancbmk\_r benchmark exhibits complex data movement patterns that challenge traditional SimPoint methods. To address this, we propose enhancing SimPoint’s BBV methodology by incorporating Memory Access Vectors (MAV), a microarchitecture-independent technique that tracks functional memory access patterns. This combined approach significantly improves the projection accuracy of 523.xalancbmk\_r on a 192-core system-on-chip, increasing it from 80\% to 98\%.

\end{abstract}


\section{Introduction}

Performance evaluation is a critical component in the design and development of modern microprocessors \cite{intel}. Architects construct software models of the CPU to simulate notable applications. In industrial settings, performance models serve four primary purposes:

\begin{enumerate}
\item \textbf{Microarchitecture Sandbox}: A modeling environment to implement, debug, and analyze new feature ideas for future CPU products.
\item \textbf{Performance Verification}: A reference implementation of performance behaviors to validate against the RTL design, identifying performance bugs.
\item \textbf{Software Projection}: A forecasting tool to estimate benchmark scores for future CPUs, informing the product team whether the design meets target specifications.
\item \textbf{Software Tuning on Future Hardware}: A virtual platform for developers to optimize software on pre-silicon hardware.
\end{enumerate}

Improvements in any of these areas enhance the others. This paper focuses on the third pillar: software projection.

Software performance projections are typically conducted using sampling techniques such as SimPoint \cite{simpoint1, simpoint2}. Workloads are preprocessed through checkpoints, and SimPoint is used to determine where to collect traces for simulation through analyzing Basic Block Vectors (BBVs)\cite{simpoint3}. Sampling is necessary because performance simulations are significantly slower than running applications on silicon. To enhance the accuracy of projections, we track the performance model's accuracy against the silicon it represents. Improvements can be made by either refining the model's accuracy or optimizing the sampling methodology for selecting which parts of the application to simulate.

One of the benchmarks we estimate is the SPEC CPU2017 integer suite \cite{bucek}. Using our in-house performance simulator, we compare the SimPoint-BBV projected scores for each component of the CPU2017 integer suite from the performance model against the AmpereOne A192-32X  \cite{amperebrief,ampere1}. The objective is to achieve a correlation as close to 1.00 as possible. As illustrated in Table~\ref{tab:projections}, the simulation model accurately estimates performance within 10\% for most benchmark/product combinations, and within 3\% on average. The big outlier is the underestimation of performance of 523.xalanc by 20\% on the AmpereOne 192-core SoC.

\begin{table}[]
    \centering
    \begin{tabular}{|l|c|c|c|}
    \hline     
       \textbf{CPU2017 benchmark}  &  \textbf{96 cores} & \textbf{128 cores} & \textbf{192 cores} \\ \hline
        500.perlbench\_r  & 0.99 & 0.98 & 0.98 \\
        502.gcc\_r        & 1.06 & 1.05 & 1.05 \\
        505.mcf\_r        & 0.88 & 0.90 & 1.03 \\
        520.omnetpp\_r    & 1.04 & 1.06 & 1.01 \\
        523.xalancbmk\_r  & \textcolor{red}{0.84} & \textcolor{red}{0.82} & \textcolor{red}{0.80} \\
        525.x264\_r       & 0.99 & 0.99 & 0.99 \\
        531.deepsjeng\_r  & 1.06 & 1.06 & 1.08 \\
        541.leela\_r      & 0.99 & 0.98 & 0.97 \\
        548.exchange2\_r  & 1.02 & 1.02 & 1.02 \\
        557.xz\_r         & 0.91 & 0.92 & 0.93 \\ \hline
    \end{tabular}
    \vspace{5pt} 
    \caption{Baseline SPECrate correlation for AmpereOne SoCs.}
    \label{tab:projections}
\end{table}

The classic sampling methodologies face limitations with workloads containing indirect memory accesses of the form \texttt{a[b[i]]}, where data access patterns significantly influence performance. Such access patterns are prevalent in graph workloads and machine learning inference applications \cite{chilu,jang,balaji}. The same code region can exhibit different microarchitectural phases depending on memory access characteristics, such as the program's working set size and access distribution.

523.xalanc exemplifies this phenomenon. While we have developed microbenchmarks to isolate this behavior and are aware of other applications exhibiting similar traits, we demonstrate it here using a well-known published benchmark. 523.xalanc's memory access patterns illustrate shifts in microarchitectural phases, observable both on silicon and in pre-silicon performance models. We introduce the concept of Memory Access Vectors (MAV) with a methodology that captures these behaviors and improves sampling accuracy.


\section{Historical Works}

SimPoint is a well-established methodology that leverages Basic Block Vectors (BBVs) to identify program phases, owing to the strong correlation between code signatures and performance characteristics \cite{lau1}. A basic block is defined as a code segment with a single entry and exit point. This methodology involves counting the occurrences of basic blocks within a specified instruction window to construct a vector. These vectors, representing different execution windows, are then compared using Euclidean or Manhattan distance measures to determine similarity scores. High similarity scores indicate that the code executions within those windows are analogous, suggesting similar microarchitectural phases, such as instructions-per-cycle (IPC) metrics and cache miss rates.

S.~Singh et al. meticulously document this process for SPEC CPU 2017 \cite{singh} and evaluate the accuracy of the methodology across its sub-components. However, their study notably omits 523.xalanc. Although 623.xalanc is included, it is recognized that the Xalan application exhibits varied behavior under different input loads \cite{alberta}. Upon inquiry, the authors clarified that 523.xalanc was excluded due to its convergence issues with the classic technique, resulting in incomplete runs.

Researchers have investigated various methods to improve the sampling fidelity of SimPoint \cite{tracedoctor}, including the incorporation of microarchitectural performance and power metrics \cite{smarts,nps}. Other studies have emphasized the value of microarchitecture-independent techniques, such as the Reuse Distance Distribution (RDD) \cite{rdd} to capture relative memory access patterns. Although RDD and MAV both focus on characterizing memory behavior, MAV presents practical advantages over RDD in certain scenarios. Unlike RDD, which was designed as an alternative to BBV, MAV was specifically developed to complement BBV through a straightforward weighting mechanism that dynamically adjusts to an application's memory intensity. Additionally, RDD's reuse calculations are computationally intensive compared to what MAV offers with frequency counts. Similarly, CompressPoints \cite{compresspoints} identifies program phases using memory compressibility patterns, though it focuses on compression efficiency of data rather than analyzing address patterns.

\section{Memory Access Vectors}

To address phase changes and enhance correlation at high core counts, we introduce the concept of Memory Access Vectors (MAV). Analogous to the BBV technique, MAV segments program execution into instruction windows. Within each window, it tracks read and write operations to unique memory blocks in the physical address space. All memory accesses are recorded based on the functional execution of the program, independent of microarchitectural caches or TLBs. Unlike the Reuse Distance Distribution (RDD) concept, MAV focuses on absolute addresses and access frequencies rather than deltas between accesses. This distinction is particularly crucial for \emph{refrate}-style homogeneous runs, as total counts are compounded with many cores running.

\begin{figure*}[t]
    \centering
    \includegraphics[width=\textwidth]{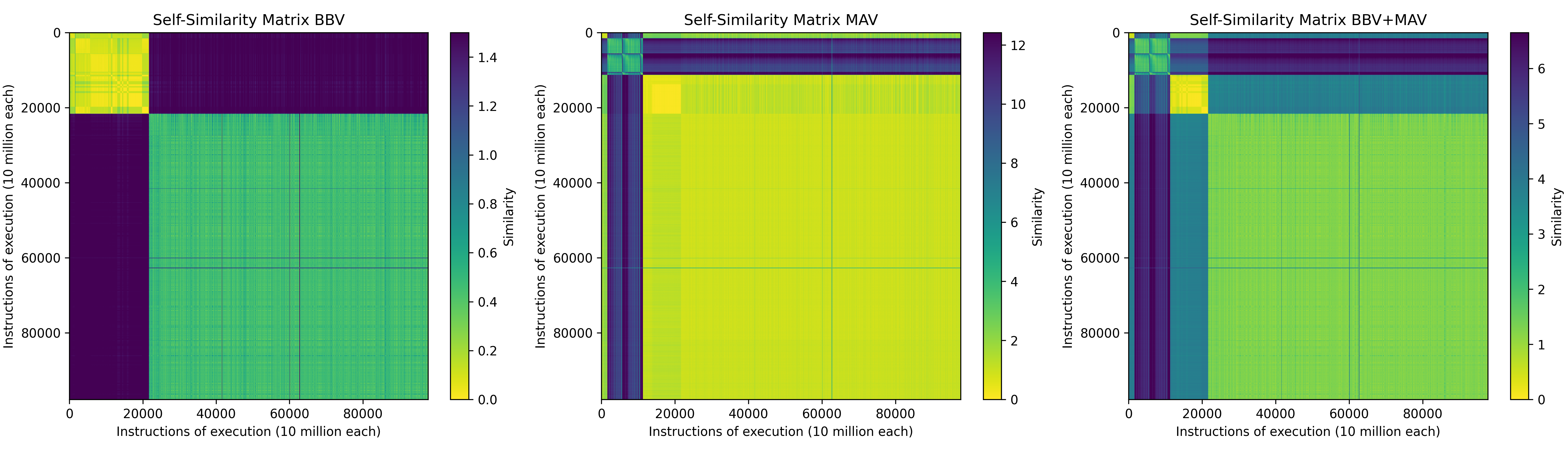}
    \caption{Self-Similarity plots of 523.xalancbmk\_r showing BBV, MAV, and combined BBV+MAV.}
    \label{fig:ss_plot}
\end{figure*}


To effectively utilize MAVs within the SimPoint framework, a processing flow has been developed that integrates with the existing BBV methodology:


\textbf{1. Vector Transformation}: When processing Memory Access Vectors (MAVs) for similarity comparison, the inverse of each memory region's access frequency is computed, and the vector is sorted in descending order of these inverse frequencies. This step emphasizes regions that are accessed infrequently, which are likely to cause cache misses or page faults, over frequently accessed regions that are likely cached. This alignment focuses the signature on actual performance impact. The memory address labels are then discarded, retaining only the ordered frequency distribution. This transformation shifts the comparison's focus to the pattern of performance-critical accesses rather than specific memory locations.

\textbf{2. Normalization}: Unlike Basic Block Vectors (BBVs), where each vector is normalized individually, the entire MAV matrix is normalized by dividing each row by the average magnitude across all rows. This approach preserves the relative intensity of memory operations across instruction windows, providing valuable information about memory pressure in different program phases.

\textbf{3. Temporal Locality}: Memory access patterns exhibit temporal locality. Given the focus on large server-class CPUs with extensive cache hierarchies, longer-term memory reuse is captured by applying a 0.95 exponential decay over the previous 10 instruction windows. This method prioritizes recent behavior while incorporating the lingering effects of prior memory access patterns.

\textbf{4. Dimension Reduction}: Gaussian Random Projection is applied to both BBV and MAV matrices, reducing each to 15 dimensions.  BBVs are already 15-dimensional by default, so this ensures that MAVs are given equal weight in terms of dimensionality. The reduced matrices are then concatenated to form a combined 30-dimensional representation for each instruction window.

\textbf{5. Adaptive Weighting}: To prevent MAVs from dominating the clustering process in code-intensive applications, the MAV contribution is scaled by multiplying its values by the percentage of memory operations in the entire application. This scaling ensures that MAVs significantly influence phase detection in memory-intensive applications, while BBVs remain the primary phase indicator in compute-bound applications with fewer memory operations. This approach allows the weighting to automatically adapt to the characteristics of each application without requiring manual tuning. Even in compute-bound applications with a high ratio of memory operations, if the data footprint is small or the access pattern is simple, the MAVs may have a high weight but will exhibit minimal variability, thus having limited influence on the final selection.

\textbf{6. Clustering}: The combined and weighted BBV+MAV matrices are then fed into SimPoint's k-means algorithm for clustering and representative selection.

This workflow implements MAV in a manner that complements existing SimPoint methodology while addressing its limitations for workloads with indirect memory access patterns. The transformation, normalization, and weighting steps ensure that both code and memory access patterns contribute appropriately to phase detection.


\section{Analysis}

\subsection{Implementation}

BBVs can be collected on silicon using Valgrind \cite{weaver1} or through emulation using QEMU \cite{weaver2}. We chose to implement MAV by instrumenting QEMU, given its support for future ISAs. We utilize an instruction length of 10M and run benchmarks through QEMU to collect data.

The MAV collection and processing involve two steps. First, a memory granularity must be specified to create the histogram buckets. Second, each access within the granular range increments the count in the corresponding bucket. The chosen granularity should allow for a sufficient number of buckets within the instruction window to distinguish different behaviors. A granularity that is too small results in large vectors that are computationally intensive to process, while a granularity that is too large fails to capture meaningful memory access patterns. We selected 4096 bytes as the granularity, as it aligns with the common memory page size in modern operating systems and empirically meets practical runtime requirements. The MAV output consists of a histogram of access counts for each 4096 byte region, with one vector per instruction epoch. The computational overhead of MAV collection is minimal, requiring only histogram updates during the QEMU data collection phase, and no impact on trace-collection phase after SimPoints have been identified.

We applied the flow above to identify 30 SimPoint clusters for the 523.xalanc benchmark, and we share our findings below.


\subsection{Recurrence plots}

Recurrence \cite{recurrence} or self-similarity is a technique employed in the analysis of music \cite{music, audio} and vision \cite{vision} to identify repeating patterns through deltas in large data matrices. We also utilize this technique to visualize recurrent behavior in programs, as observed through the distances between individual vectors of basic blocks and memory accesses.

The plots in Figure~\ref{fig:ss_plot} illustrate recurrence in the 980 billion instructions of 523.xalanc, segmented into chunks of 10 million instructions each. The left image presents the traditional BBV plot, indicating code similarity within the first 200 billion instructions, which corresponds to the section of the benchmark executing the Xerces-C++ 2.7 parser. The subsequent 700 billion instructions, running the Xalan-C++ 1.1 transformer, exhibit more varied behavior. The center image introduces the new MAV plot, highlighting data similarity between 100 billion and 200 billion instructions, with increased similarity from 100 billion instructions to the end of the program. 

The discrepancy between these first two images enables us to isolate the problematic region: the parser accesses a variety of distinct data regions, despite the recurring code. Subsequently, the parser is passing its processed data to the transformer. Our technique combines BBV and MAV, weighted according to the percentage of memory instructions, to produce the image on the right. This combined approach reveals multiple phases within the first 200 billion instructions, which are not discernible using BBV or MAV alone.


\subsection{Phase plots}

This analysis employs 30 clusters, a common choice for benchmarks of this complexity level. The integration of BBV and MAV into the methodology does not necessitate alterations to the cluster selection process; established SimPoint heuristics for determining optimal cluster counts remain valid. 

\begin{figure}[h]
    \centering
        \includegraphics[width=\columnwidth]{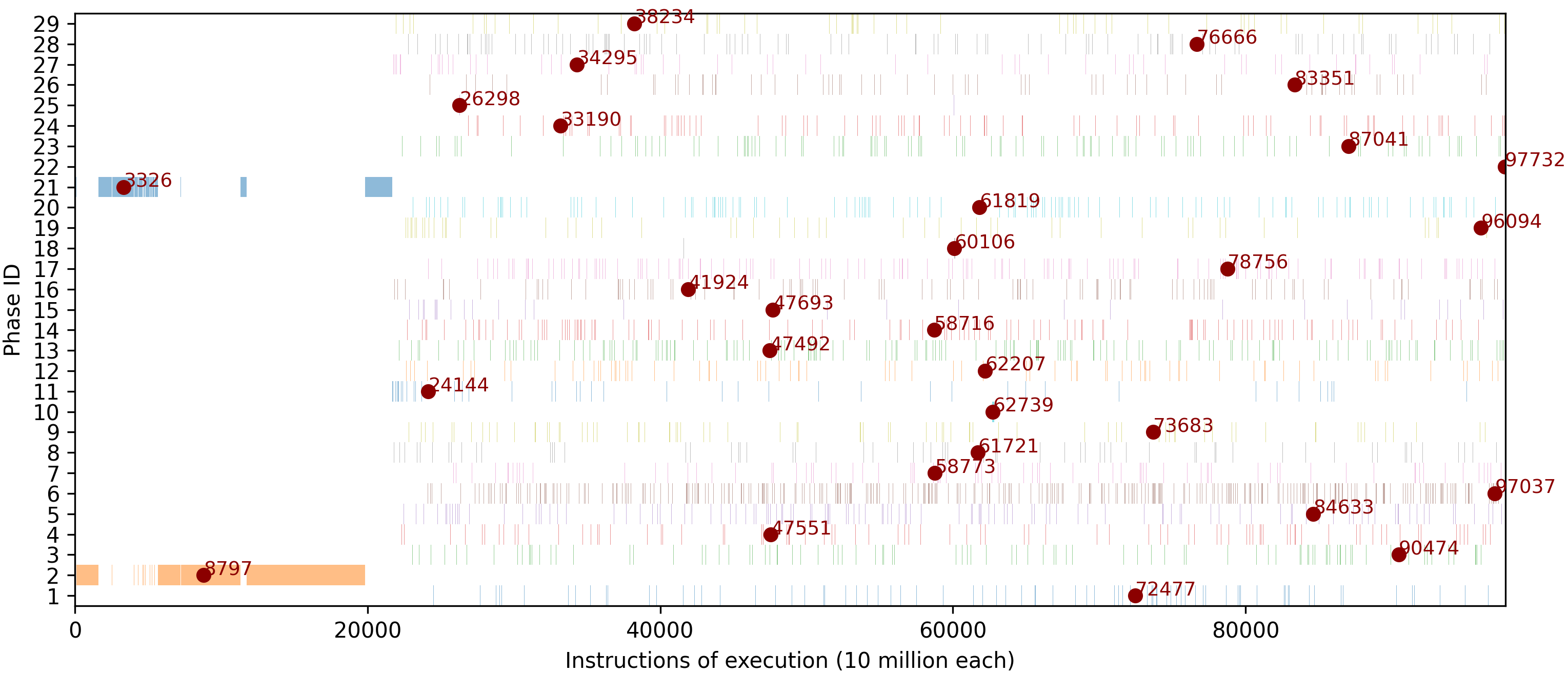}
        \caption{BBV-only phases and SimPoint selections for 523.xalanc.}
        \label{fig:bbv_only}
        \vfill{\ }
        \includegraphics[width=\columnwidth]{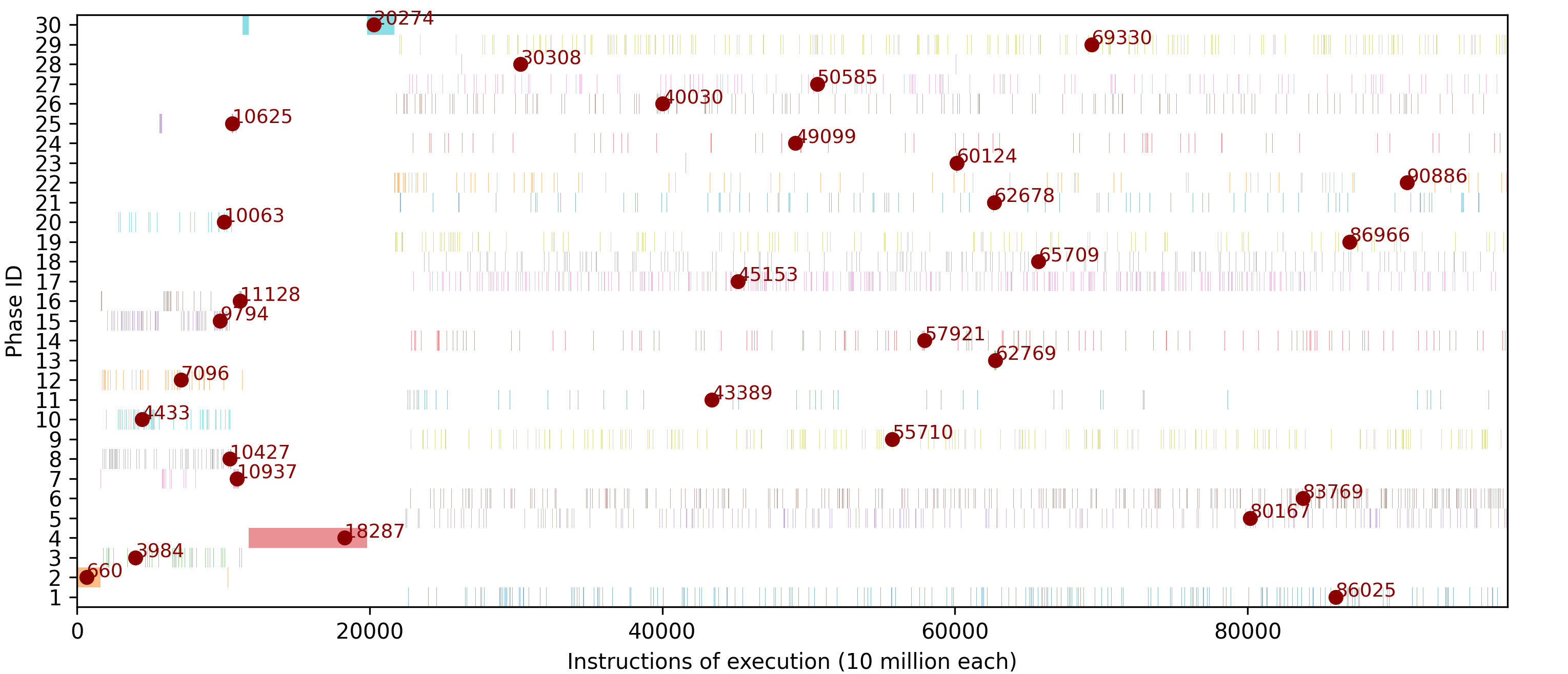}
        \caption{BBV+MAV phases and SimPoint selections for 523.xalanc.}
        \label{fig:bbv_mav}
        \vfill{\ }
        \includegraphics[width=1.01\columnwidth]{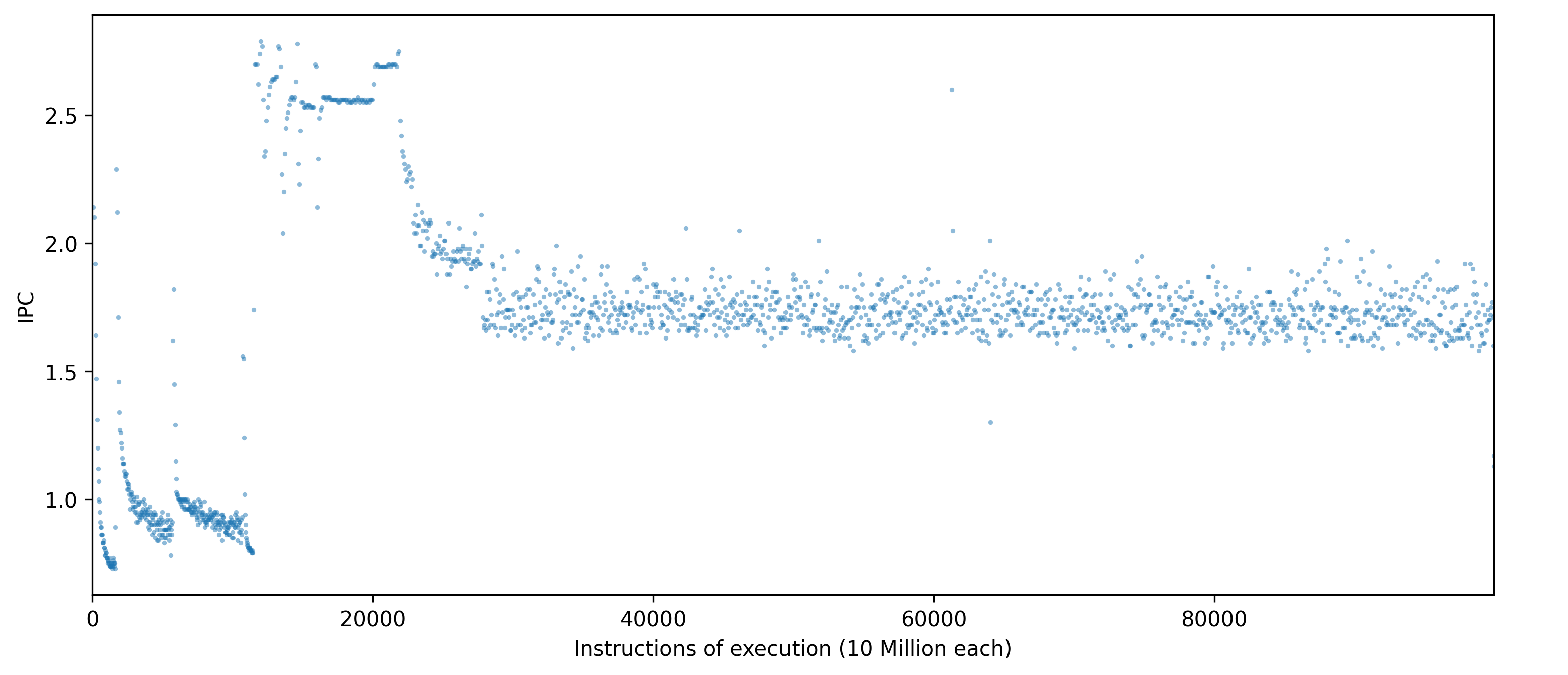}
        \caption{IPC plot of 523.xalanc on AmpereOne silicon.}
        \label{fig:ipc_plot}
\end{figure}

The baseline phase plot in Figure~\ref{fig:bbv_only} provides a detailed examination of how SimPoint identifies 30 phases and the decisions it makes for k-means clustering. Using BBV alone, only two phases cover the first 200 billion instructions (Phase IDs 2 and 21). This indicates that BBV by itself considers this region to be homogeneous and requires only two samples\footnote{The two hot methods in Xerces are \texttt{ValueStore::isDuplicateOf} and \texttt{ValueStore::contains}.}.

Figure~\ref{fig:bbv_mav} illustrates the changes resulting from the combination of BBV and MAV. SimPoint is now able to discern a clearer separation for clustering, and increases the sample points for the Xerces region to 12, constituting over a third of the total clusters. Given the fixed number of clusters, the samples shift away from the Xalan portion of the benchmark, which is now perceived as less diverse. Consequently, the Xerces portion requires more samples to adequately cover all behaviors. Despite the presence of only two hot methods, the sample selections count increases, driven by the nature of varying data being processed.

Alongside these two phase plots, Figure~\ref{fig:ipc_plot} displays the instructions-per-cycle (IPC) of each 10 million instruction region on AmpereOne silicon. This plot allows us to verify whether SimPoint, using the combined BBV+MAV approach, effectively clusters to represent microarchitectural behavior on a first-order basis. Overlaying Figure~\ref{fig:bbv_only} with the IPC plot reveals that Phase 2 was attempting to cover regions with both very low and very high IPC, which is clearly inadequate. In contrast, comparing Figure~\ref{fig:bbv_mav} with the IPC plot shows that the complex IPC behaviors of the Xerces region are now represented by multiple samples, with the highest IPC regions represented by BBV+MAV Phases 4 and 30.

\subsection{Results}

Using the new SimPoints shown in Figure~\ref{fig:bbv_mav}, we have rerun our performance projections through our standard flow. The result of using these more accurate SimPoints is an improvement in correlation numbers, bringing the software projection for the simulation model closer to the silicon's reported performance. Table~\ref{tab:projections2} presents the before-and-after comparison for the outlier benchmark, demonstrating that the correlation for the 192-core SoC improved from 80\% to 98\%. 

\begin{table}[t]
    \centering
    \begin{tabular}{|l|c|c|c|}
    \hline     
       \textbf{sampling technique}  &  \textbf{96 cores} & \textbf{192 cores} \\ \hline
        523.xalancbmk\_r: BBV only & 0.84 & 0.80 \\
        523.xalancbmk\_r: BBV+MAV & 0.95 & 0.98 \\  \hline
    \end{tabular}
    \vspace{5pt} 
    \caption{Correlation of both sampling techniques on AmpereOne SoCs.}
    \label{tab:projections2}
\end{table}


\section{Conclusion and Future Work}

By refining the sampling methodology to add memory access information, we have increased confidence in the projections our simulations provide for future products. This new approach is invaluable for projecting the performance of applications that exhibit extensive array-indirect memory accesses and has the potential to enhance the accuracy of projections for a wide range of benchmarks. As an example, we showed a correlation improvement on 523.xalancbmk\_r, bringing the estimation within 2\% of the 192-core silicon measurement. Our future work will formalize this sampling process and explore its applicability to more applications as well as even larger core counts.




\begin{thebibliography}{9}

\bibitem{amperebrief}
AmpereOne{\textregistered} Product Brief (A192-32X), Ampere Computing. [Online] \url{https://amperecomputing.com/briefs/ampereone-family-product-brief}

\bibitem{ampere1}
D.~Carlson, N.~Simakov, R.~Hadlich, A.~Curtis, J.~Martin, G.~Verma, S.~Chheda, F.~Coskun, R.~Gonzalez, D.~Wood, F.~Zhang, R.~Harrison, E.~Siegmann, \emph{The AmpereOne A192-32X in Perspective: Benchmarking a New Standard}. In Proceedings of the 2025 International Conference on High Performance Computing in Asia-Pacific Region Workshops. Association for Computing Machinery, New York, NY, USA. Available: \url{https://doi.org/10.1145/3703001.3724384}

\bibitem{bucek}
J.~Bucek, K-D.~Lange, J.~Kistowski, \emph{SPEC CPU2017: Next-Generation Compute Benchmark}, In Companion of the 2018 ACM/SPEC International Conference on Performance Engineering (ICPE '18). Association for Computing Machinery, New York, NY, USA. Available: \url{https://doi.org/10.1145/3185768.3185771}

\bibitem{simpoint1}
T.~Sherwood, E.~Perelman, G.~Hamerly, B.~Calder, \emph{Automatically characterizing large scale program behavior}, ACM International Conference on on Architectural Support for Programming Languages and Operating Systems, October 2002. Available: 
  \url{https://api.semanticscholar.org/CorpusID:6445892}

\bibitem{simpoint2}
G.~Hamerly, E.~Perelman, J.~Lau, B.~Calder, \emph{SimPoint 3.0: Faster and More Flexible Program Phase Analysis}, Journal of Instruction-Level Parallelism, Volume 7, September 2005. Available: \url{https://api.semanticscholar.org/CorpusID:11937761}

\bibitem{simpoint3}
B.~Calder, T.~Sherwood, G.~Hamerly, E.~Perelman, \emph{SimPoint: Picking Representative Samples to Guide
Simulation}, book chapter from \emph{Processor and System-on-Chip Simulation} edited by R.~Leupers and O.~Temam,  Springer ISBN 978-1-4419-6174-7, 2010. Available: \url{https://link.springer.com/book/10.1007/978-1-4419-6175-4} 

\bibitem{singh}
S.~Singh and M.~Awasthi, \emph{Efficacy of Statistical Sampling on Contemporary Workloads: The Case of SPEC CPU2017}, IEEE International Symposium on Workload Characterization (IISWC), Orlando, FL, USA, 2019, doi: 10.1109/IISWC47752.2019.9042114. Available: \url{https://ieeexplore.ieee.org/document/9042114}

\bibitem{alberta}
E.~Colp, J.N.~Amaral, C.~Benedicto, R.E.~Rodrigues, E.~Borin, \emph{Report: The 523.xalancbmk\_r Benchmark}, University of Alberta, January 2018. [Online] Available: \url{https://webdocs.cs.ualberta.ca/~amaral/AlbertaWorkloadsForSPECCPU2017/reports/xalancbmk_report.html}

\bibitem{lau1}
J.~Lau, J.~Sampson, E.~Perelman, G.~Hamerly, B.~Calder, \emph{The strong correlation between code signatures
and performance}, International Symposium on Performance Analysis of Systems and Software, March 2005. Available: \url{https://ieeexplore.ieee.org/document/1430578}

\bibitem{intel}
R.~Singhal, et.al, \emph{Performance Analysis and Validation of the Intel {\textregistered} Pentium {\textregistered} 4 Processor on 90 nm Technology}, Intel Technology Journal, 2004. Available: \url{https://api.semanticscholar.org/CorpusID:17764788}

\bibitem{chilu}
A.~Chilukuri, J.~Milthorpe, B.~Johnston, \emph{Characterizing Optimizations to Memory Access Patterns using Architecture-Independent Program Features}, Proceedings of the International Workshop on OpenCL, 2020. Available: \url{https://doi.org/10.48550/arXiv.2003.06064}

\bibitem{jang}
J.~Jang, H.~Kim and H.~Lee, \emph{Characterizing Memory Access Patterns of Various Convolutional Neural Networks for Utilizing Processing-in-Memory}, 2023 International Conference on Electronics, Information, and Communication (ICEIC), Singapore, 2023. Available: \url{https://doi.org/10.1109/ICEIC57457.2023.10049894}

\bibitem{balaji}
V.~Balaji, N.~Crago, A.~Jaleel and B.~Lucia, \emph{P-OPT: Practical Optimal Cache Replacement for Graph Analytics}, 2021 IEEE International Symposium on High-Performance Computer Architecture (HPCA), Seoul, Korea (South), 2021. Available: \url{https://doi.org/10.1109/HPCA51647.2021.00062}

\bibitem{weaver1}
V.~Weaver, \emph{exp-bbv: Valgrind plugin that makes SimPoint Basic Block Vector Files}, [Online] \url{https://valgrind.org/docs/manual/bbv-manual.html} and \url{https://web.eece.maine.edu/~vweaver/projects/valsim}

\bibitem{weaver2}
V.~Weaver, \emph{qemu\_bbv - a qemu patch that enables SimPoint Basic Block Vector File Generation}, [Online] \url{https://web.eece.maine.edu/~vweaver/projects/qemusim}

\bibitem{tracedoctor}
B.~Gottschall, S.~C.~de~Santana, M.~Jahre, \emph{Balancing Accuracy and Evaluation Overhead in Simulation Point Selection}, 2023 IEEE International Symposium on Workload Characterization (IISWC), Ghent, Belgium, 2023, pp. 43-53, doi: 10.1109/IISWC59245.2023.00019. Available: \url{https://doi.org/10.1109/IISWC59245.2023.00019}

\bibitem{smarts}
R.~E.~Wunderlich, T.~F.~Wenisch, B.~Falsafi, J.~C.~Hoe, \emph{SMARTS: accelerating microarchitecture simulation via rigorous statistical sampling}, 30th Annual International Symposium on Computer Architecture, 2003. San Diego, CA, USA, 2003, pp. 84-95, doi: 10.1109/ISCA.2003.1206991. Available: \url{https://ieeexplore.ieee.org/document/1206991}

\bibitem{nps}
Y.~Fang, et.al, \emph{NPS: A Framework for Accurate Program Sampling Using Graph Neural Network}, arXiv:2304.08880. April 2023. Available: \url{https://arxiv.org/abs/2304.08880}

\bibitem{rdd}
Y.~Luo, A.~Joshi, A.~Phansalkar, L.~John, J.~Ghosh, \emph{Analyzing and improving clustering based sampling for microprocessor simulation}, 17th International Symposium on Computer Architecture and High Performance Computing (SBAC-PAD'05), Rio de Janeiro, Brazil, 2005. doi: 10.1109/CAHPC.2005.11. Available: \url{https://ieeexplore.ieee.org/document/1592573}

\bibitem{compresspoints}
E.~Choukse, M.~Erez,  A.~Alameldeen, \emph{CompressPoints: An Evaluation Methodology for Compressed Memory Systems}, IEEE Computer Architecture Letters, vol. 17, no. 2, pp. 126-129, 2018. Available: \url{https://ieeexplore.ieee.org/document/8328832}

\bibitem{recurrence}
N.~Marwan, M.C.~Romano, M.~Thiel, J.~Kurths, \emph{Recurrence plots for the analysis of complex systems}, Physics Reports, Volume 438, Issues 5–6, January 2007, Pages 237-329, ISSN 0370-1573, arXiv:2501.13933. Available: \url{https://doi.org/10.1016/j.physrep.2006.11.001} and \url{https://doi.org/10.48550/arXiv.2501.13933}

\bibitem{music}
J.~Foote, \emph{Visualizing music and audio using self-similarity}, In Proceedings of the seventh ACM international conference on Multimedia (Part 1) (MULTIMEDIA '99). Association for Computing Machinery, New York, NY, USA, 77–80. Available: \url{https://doi.org/10.1145/319463.319472}

\bibitem{audio}
M.~Müller, M.~Clausen, \emph{Transposition-invariant self-similarity matrices}. Proceedings of the 8th International Conference on Music Information Retrieval (ISMIR 2007): 47–50. Available: \url{https://www.researchgate.net/publication/220723240_Transposition-Invariant_Self-Similarity_Matrices}

\bibitem{vision}
I.N.~Junejo, E.~Dexter, I.~Laptev, P.~Pérez, \emph{Cross-View Action Recognition from Temporal Self-similarities}. In: Computer Vision – ECCV 2008. Lecture Notes in Computer Science, vol 5303. Springer, Berlin, Heidelberg. ISBN 978-3-540-88685-3. Available: \url{https://doi.org/10.1007/978-3-540-88688-4_22}

\end{thebibliography}


\end{document}